\documentclass[aps,pra,twocolumn,groupaddress]{revtex4}
\usepackage{graphicx}
\usepackage{hyperref}
\usepackage{amsmath,amssymb}
\usepackage{bm}
\begin{document}
\title{Spin transport in a unitary Fermi gas close to the BCS transition}
\author{M. P. Mink}
\email{m.p.mink@uu.nl}
\affiliation{Institute for Theoretical Physics, Utrecht
University, Leuvenlaan 4, 3584 CE Utrecht, The Netherlands}
\author{V. P. J. Jacobs}
\affiliation{Institute for Theoretical Physics, Utrecht
University, Leuvenlaan 4, 3584 CE Utrecht, The Netherlands}
\author{H. T. C. Stoof}
\affiliation{Institute for Theoretical Physics, Utrecht
University, Leuvenlaan 4, 3584 CE Utrecht, The Netherlands}
\author{R. A. Duine}
\affiliation{Institute for Theoretical Physics, Utrecht
University, Leuvenlaan 4, 3584 CE Utrecht, The Netherlands}
\author{Marco Polini}
\affiliation{NEST, Istituto Nanoscienze-CNR and Scuola Normale Superiore, I-56126 Pisa, Italy}
\author{G. Vignale}
\affiliation{Department of Physics and Astronomy, University of Missouri, Columbia, Missouri 65211, USA}
\date{\today}
\begin{abstract}
We consider spin transport in a two-component ultracold Fermi gas with attractive interspecies interactions close to the BCS pairing transition. In particular, we consider the spin-transport relaxation rate and the spin-diffusion constant. Upon approaching the transition, the scattering amplitude is enhanced by pairing fluctuations. However, as the system approaches the transition, the spectral weight for excitations close to the Fermi level is decreased by the formation of a pseudogap. To study the consequence of these two competing effects, we determine the spin-transport relaxation rate and the spin-diffusion constant using both a Boltzmann approach and a diagrammatic approach. The former ignores pseudogap physics and finite lifetime effects. In the latter, we incorporate the full pseudogap physics and lifetime effects, but we ignore vertex corrections, so that we effectively calculate single-particle relaxation rates instead of transport relaxation rates. We find that there is qualitative agreement between these two approaches although the results for the transport coefficients differ quantitatively.
\end{abstract}
\maketitle
\vskip2pc
\section{Introduction}
\label{sec:int}
Over the last decade, there has been a growing interest in the physics community in the properties of spin transport, spurred by the idea of using the electron spin as a carrier of information. A spin current is a net flow of spin and is fundamentally different from a charge current, because it relaxes in a different way. In particular, spin currents can be strongly affected by interactions, even in Galilean-invariant systems.

Ultracold fermion systems consisting of two spin species provide a valuable model system for the study of the effects of interactions on spin transport, because of their high tunability and the absence of other factors which limit the spin conductivity, such as disorder. When the cloud of one species moves relatively to the cloud of the other species, the latter is dragged along due to momentum relaxation by the interatomic interaction, and as a consequence, the spin current relaxes. This mechanism is called spin drag \cite{spindrag}. A similar phenomenon, Coulomb drag, occurs in bilayer systems, in which the drift momentum difference between the carriers in the top and bottom layer relaxes due to the Coulomb interaction \cite{gramilla,rojo}.

In an important recent experimental study, the spin susceptibility, the spin-diffusion constant, and the relaxation rate of spin transport were investigated in a ultracold two-component fermion gas in the unitarity regime, where the interspecies scattering length goes to infinity so that the interactions are as strong as quantum mechanics allows \cite{sommer,meta}. Inspired by this work, the spin-transport relaxation rate and the spin-diffusion constant were calculated using a Boltzmann approach for high and low temperature ranges separately \cite{bruun2011}.

When an ultracold Fermi gas with two spin species and an attractive interaction between the spins is cooled to low enough temperatures it shows a transition to a superfluid state, where the opposite-spin atoms pair up to form Cooper pairs. The effect of interactions on spin transport close to this BCS pairing transition is an interesting subject, since two effects are competing. On the one hand, the scattering amplitude between fermions is enhanced by pairing fluctuations (preformed Cooper pairs) not taken into account in Ref.~\cite{bruun2011}. When the temperature is lowered, this effect ultimately leads to a diverging interaction strength and an instability in the system towards the BCS state at the critical temperature. As a consequence, transport coefficients like the spin-transport relaxation rate and the spin-diffusion constant, are expected to be strongly affected when the system approaches the transition. These effects were not seen experimentally in Ref.~\cite{sommer}, possibly because $T_c$ was not reached but possibly also due to the competing effect we discuss shortly.

In related work, for ultracold fermions with repulsive interactions, enhancement of the spin-transport relaxation rate was predicted close to the ferromagnetic transition \cite{duinemag}, and for ultracold bosons close to Bose-Einstein condensation \cite{hedwig}. The latter was recently also observed experimentally \cite{straten}. Riedl {\it et al.} studied the frequencies and damping of collective modes in a ultracold Fermi gas close to the BCS pairing transition \cite{riedl} and found an enhancement of the collision rate in a fermionic gas close to the BCS pairing transition. The behavior of the viscosity was studied in Ref.~\cite{bruun2009}. In electron-hole bilayers the electrons in one layer and holes in the other can form excitons, which are expected to condense for low enough temperatures. Theoretically, it was predicted that the transport relaxation rate is enhanced when approaching this transition \cite{hu}. An enhancement was measured experimentally \cite{ehexp}, although it is still under debate whether this enhancement was indeed caused by exciton condensation. Also for a topological insulator thin film, an enhancement of the transport relaxation rate was predicted close to (in this case topological) exciton condensation \cite{mink}.

However, as already mentioned, there is a competing effect at play. For temperatures below the transition temperature, the system is in the superfluid state and the excitation spectrum is gapped. Already above $T_c$, a precursor of this gap can be seen as the suppression of spectral weight close to the Fermi level, a so-called pseudogap \cite{pseud}. The suppressed spectral weight is closely linked to the reduced lifetime for these excitations. The effect of this pseudogap is to reduce the phase space for scattering events around the Fermi level, which is expected to reduce the enhancement of transport coefficients. It is the competition between the increase in scattering amplitude due to pairing fluctuations on the one hand, and the decrease in available phase space due to the pseudogap on the other, and its effect on the transport relaxation rate and the spin-diffusion constant that we consider in this paper. This competition was considered in Ref.~\cite{theosuscep} for the spin susceptibility. We note related work determining the static spin susceptibility for repulsive interactions in Ref.~\cite{string}.

In general, the relaxation rate of a current and the decay rate of a single-particle excitation are different because of the amount with which the possible scattering directions contribute to either case. In the case of the decay of a single-particle excitation, all directions are weighed equally, but in the case of current relaxation, forward scattering (in the direction of the current) contributes much less than back scattering (in the direction opposite to the current). Theoretically, in diagrammatic calculations, a transport relaxation rate is obtained by inclusion of so-called vertex corrections to the appropriate response function. If these are neglected, the transport relaxation time is essentially equal to the single-particle relaxation time.

In this paper we consider a two-component gas of fermions consisting of two spin states labeled by $\sigma = \uparrow,\downarrow$. We consider the balanced case where the densities of each spin component are equal $n_\uparrow = n_\downarrow \equiv n$, and where both species have the dispersion $\xi({\bm k}) = \hbar^2 k^2/2m - \mu$. This article consists of two parts. In the first part in Sec.~\ref{sec:BM} we use Boltzmann theory to calculate the relaxation rate for spin transport in this system for arbitrary interaction strength, both close to the BCS pairing transition and for temperatures much higher than the Fermi temperature. Then, using the non-interacting spin susceptibility, we can determine the spin-diffusion constant using the Einstein relation. This calculation incorporates the effect of pairing fluctuations, but does not take into account the pseudogap physics or the effect of the finite lifetime of the quasiparticle eigenstates. At the diagrammatic level, however, this calculation includes vertex corrections.

In the second part in Sec.~\ref{sec:dia} we calculate the fermion self-energy at unitarity both close to the BCS pairing transition and for high temperatures within the many-body $T$-matrix approximation. This self-energy is the many-body $T$-matrix closed with a bare (non-interacting) fermion line. Using this self-energy, we determine the spectral function, and the spin-transport relaxation rate and spin susceptibility ignoring vertex corrections. Again, the spin-diffusion constant can be obtained using the Einstein relation. We end in Sec.~\ref{sec:con} with our conclusions.

By comparing the spin-transport relaxation rate and diffusion constant obtained using both methods, we assess the importance of pseudogap physics, finite lifetime effects, and vertex corrections. We find that although the results of these two methods for the transport coefficients differ quantitatively, there is qualitative agreement. This indicates that pseudogap physics, finite lifetime effects, and vertex corrections do not have a critical influence on the prediction using Boltzmann theory for the behavior of transport coefficients close to the BCS transition.
\section{Boltzmann Theory}
\label{sec:BM}
In this section we derive an expression for the spin-transport relaxation rate using the Boltzmann equation. We incorporate many-body effects, so-called pair correlations, close the BCS pairing transition.  We apply a spin dependent driving force ${\bm F}_\uparrow = {\bm F}$ and ${\bm F}_\downarrow = -{\bm F}$ which will give rise to a spin current ${\bm j}_s$. The momentum increase by the driving force is balanced by the momentum relaxation due to the interaction between opposite spins, so that the system reaches a steady state. This mechanism is called spin drag. In linear response, the quantum kinetic equations are written as
\begin{align}
\frac{1}{\hbar} {\bm F} \cdot \partial_{\bm k} n_F(\xi({\bm k}))     & = \Gamma_{\uparrow}({\bm k}) \label{eq:BM1} \\
-\frac{1}{\hbar} {\bm F} \cdot \partial_{\bm k} n_F(\xi({\bm k}))  & = \Gamma_{\downarrow}({\bm k}) \label{eq:BM2},
\end{align}
where $n_F(\epsilon) = 1/[1+\exp( \beta \epsilon)]$ is the Fermi-Dirac distribution with $\beta = 1 / k_B T$ the inverse thermal energy and where the collision integral $\Gamma_{\sigma}({\bm k})$ gives the net particle flux into the $({\bm k},|\sigma\rangle)$ state due to interspecies interactions. The Fermi's golden rule expression for $\Gamma_{\uparrow}({\bm k})$ is
\begin{multline} \label{eq:colint}
\Gamma_\uparrow({\bm k}) = \frac{2 \pi}{\hbar V^2}  \sum_{{\bm k}_1,{\bm k}_2,{\bm k}_3,{\bm k}_4}
\delta_{{\bm k}_1 + {\bm k}_2,{\bm k}_3 + {\bm k}_4} \delta(\xi_1 + \xi_2 - \xi_3 - \xi_4) \\ \times |T^{MB}|^2 f_{\uparrow,1} f_{\downarrow,2} (1-f_{\downarrow,3}) (1- f_{\uparrow,4}) (\delta_{{\bm k}_1,{\bm k}} - \delta_{{\bm k}_4,{\bm k}}),
\end{multline}
where $V$ is the volume and where we used a shorthand notation for the energy $\xi_i = \xi({\bm k}_i)$ and for the non-equilibrium distributions $f_{\sigma,i} = f_{\sigma}({\bm k}_i)$. We note that an analogous expression holds for $\Gamma_\downarrow({\bm k})$. Eq.~(\ref{eq:colint}) considers the total effect of incoming particles with momenta and spin ${\bm k}_1,\uparrow$ and ${\bm k}_2,\downarrow$ that are scattered into momenta and spin ${\bm k}_3,\downarrow$ and ${\bm k}_4,\uparrow$. Momentum and energy conservation is ensured by the Kronecker and Dirac delta's, respectively. The many-body $T$-matrix $T^{MB}$ gives the scattering amplitude incorporating the effects of the medium, and will be specified below.

The distribution functions are shifted from the equilibrium distribution by a momentum ${\bm k}_\sigma = m {\bm v}_\sigma /\hbar$, where ${\bm v}$ is the so-called drift velocity, which is related to the spin current by ${\bm j}_s = n_\uparrow {\bm v}_\uparrow - n_\downarrow {\bm v}_\downarrow = 2 n {\bm v}$. Here, we used that due to symmetry, the drift velocities of the two species will be opposite and of equal magnitude for both species. The distribution function $f_{\sigma}({\bm k})$ to first order in the drift velocity is $f_{\sigma}({\bm k})= n_F(\xi_{\sigma}({\bm k})) + f^1_{\sigma}({\bm k})$, where $f^1_{\sigma}({\bm k}) =  -\hbar ({\bm v}_\sigma \cdot  {\bm k}) n_F'(\xi_{\sigma}({\bm k}))$. By subtracting Eq.~(\ref{eq:BM2}) from Eq.~(\ref{eq:BM1}), multiplying with $\hbar {\bm k}$, and summing over ${\bm k}$, we arrive at the momentum balance equation
\begin{equation} \label{eq:mombal}
 - n {\bm F} = \frac{1}{V} \sum_{\bm k} (\hbar {\bm k}) \Gamma({\bm k}) = \tilde{\Gamma} {\bm v} +\mathcal{O}(v^2),
 \end{equation}
where the coefficient $\tilde{\Gamma}$ of the linear term in the right-hand side is given by
\begin{multline} \label{eq:Gat}
\tilde{\Gamma}  = - \frac{\pi \beta \hbar}{6 V^3} \sum_{{\bm k}_i} \delta_{{\bm k}_1 + {\bm k}_2,{\bm k}_3+{\bm k}_4} \delta(\xi_1 + \xi_2 - \xi_3 - \xi_4) \\ \times |T^{MB}|^2 n_1 n_2 (1-n_3) (1- n_4)({\bm k}_1 - {\bm k}_2 + {\bm k}_3 -{\bm k}_4)^2
\end{multline}
and we introduced the shorthand notation  $n_i = n_F(\xi({\bm k}_i))$. We introduce the spin drag conductivity $\sigma_D$ via ${\bm j}_s = \sigma_D {\bm F}$ and from Eq.~(\ref{eq:mombal}) we identify $\sigma_D = - 2 n^2 / \tilde{\Gamma}$. The spin drag conductivity can be related to the spin-transport relaxation rate $1/\tau_D$ by the Drude formula $\sigma_D = 2 n \tau_D/m$ leading to $1/\tau_D = - \tilde{\Gamma}/m n$.

To make further progress, we rewrite Eq.~(\ref{eq:Gat}) in a form that is convenient to account for a divergence in the pairing channel due to pairing fluctuations close to the superfluid transition. First, we introduce an additional energy integral over the variable $\hbar \omega = \xi({\bm k}_1) + \xi({\bm k}_2)$, i.e., the sum of the energies of the two incoming particles in the scattering event. This choice is convenient because close to the superfluid transition, in the on-shell approximation, the dominant energy on which $T^{MB}$ depends is $\hbar \omega$. We stress that this choice is different from the conventional one, where $\hbar \omega$ is taken to be the difference between the energies of the incoming and outgoing $\uparrow$-particles \cite{rojo}. Next, we resolve the momentum conserving Kronecker delta $\delta_{{\bm k}_1 + {\bm k}_2,{\bm k}_3+{\bm k}_4}$ in Eq.~(\ref{eq:Gat}) by choosing ${\bm k}_1 = {\bm k} +{\bm K}/2$, ${\bm k}_2 = {\bm K}/2 - {\bm k}$, ${\bm k}_3 = {\bm k}'+{\bm K}/2$, and
${\bm k}_4 = {\bm K}/2 - {\bm k}'$ and introduce an additional energy integral via
\begin{equation} \label{eq:deltas}
\delta(\xi_1 + \xi_2 - \xi_3 - \xi_4) = \int d (\hbar \omega) \delta(\xi_1 + \xi_2 - \hbar \omega) \delta(\xi_3 + \xi_4 - \hbar\omega).
\end{equation}
Then, after using the Dirac identity $\Im[1/(x+ i 0)] = - \pi \delta(x)$, we rewrite Eq.~(\ref{eq:Gat}) into
\begin{align} \label{eq:tauinv}
\frac{1}{\tau_D} ={}&  \frac{\pi \beta \hbar}{6 \pi^2 m n} \int \frac{d{\bm k}}{(2\pi)^3}   \frac{d{\bm k}'}{(2\pi)^3}  \frac{d {\bm K}}{(2\pi)^3} d (\hbar \omega) ({\bm k}'+{\bm k})^2  \nonumber\\
&\times \frac{|T^{MB}({\bm K},\hbar \omega)|^2}{\sinh^2(\beta \hbar \omega/2)}\nonumber \\
&\times\Im\left[\frac{1-n_F(\xi({\bm k} +{\bm K}/2))- n_F(\xi({\bm K}/2 - {\bm k}))}{\hbar \omega + i 0  - (\xi({\bm k} +{\bm K}/2) + \xi({\bm K}/2 - {\bm k})) }\right]\nonumber\\
&\times\Im\left[\frac{1-n_F(\xi({\bm k}'+{\bm K}/2))-n_F(\xi({\bm K}/2 - {\bm k}'))}{\hbar \omega  + i 0  - (\xi({\bm k}'+{\bm K}/2) + \xi({\bm K}/2 - {\bm k}'))}\right].
\end{align}
The effective interaction is given by the many-body matrix element $T^{MB}({\bm K},\hbar \omega)$ which takes a simple form close to the superfluid transition in these variables
\begin{equation} \label{eq:tmb}
T^{MB}({\bm K},\hbar \omega) = \frac{V_0}{1 - V_0 \Xi({\bm K},\hbar\omega)},
\end{equation}
where $V_0 = 4 \pi a \hbar^2/m$ with $a$ the $s$-wave scattering length. The bare pairing susceptibility $\Xi$ is given by
\begin{multline} \label{eq:xi}
\Xi({\bm K},\hbar\omega) =  \frac{1}{V} \sum_{{\bm k}} \\  \left[\frac{1-n_F(\xi({\bm k}+{\bm K}/2))- n_F(\xi({\bm k}-{\bm K}/2))}{\hbar \omega + i 0  - (\xi({\bm k}+{\bm K}/2) + \xi({\bm k}-{\bm K}/2))} + \frac{1}{2 \epsilon({\bm k})} \right],
\end{multline}
with $\epsilon({\bm k}) = \hbar^2 k^2/2m$. The second term of the summand cures the ultraviolet divergence of summing the first term. Its derivation can be found in Ref.~\cite{Tmatref}. We note that the transition temperature $T_c$ to the superfluid state is determined by $1/T^{MB}(\mathbf{0},0)=0$. In Eq.~(\ref{eq:tauinv}), we can perform the integrals over ${\bm k}$, ${\bm k}'$, and the angles of ${\bm K}$ exactly. The integrals over $\omega$ and the length of ${\bm K}$ must then be performed numerically. Our theory does not take into account the superfluid gap or the presence of Cooper pairs for $T < T_c$, and is thus only valid for temperature higher than the transition temperature.

\begin{figure}
\begin{center}
\includegraphics[width = 0.45 \textwidth]{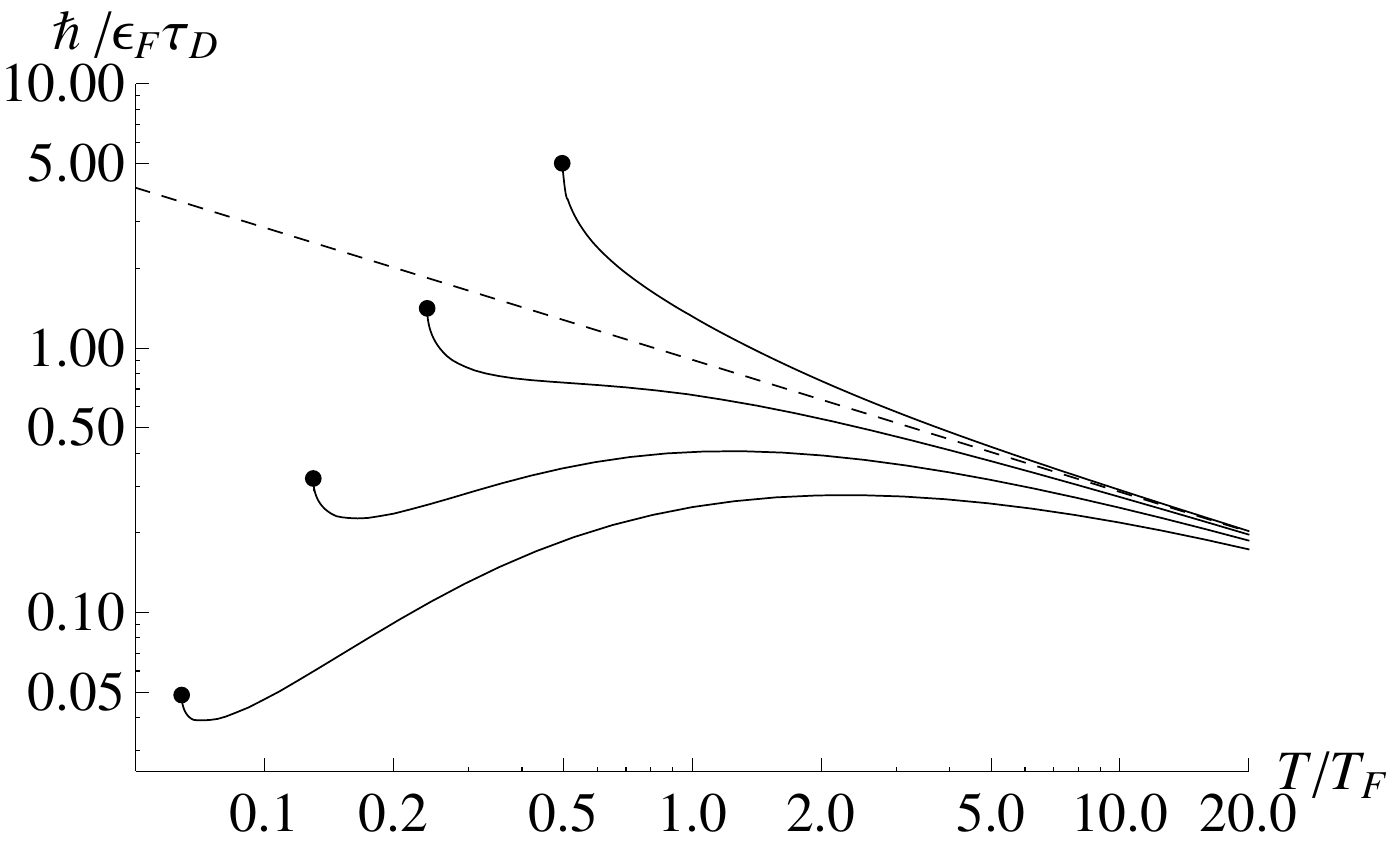}
\caption{\label{fig:BM} The dimensionless spin-transport relaxation rate $\hbar/\epsilon_F \tau_D$ obtained from Eq.~(\ref{eq:tauinv}) as a function of the reduced temperature $T/T_F$. From top to bottom the curves correspond to interaction parameters $-1/k_F a = \{0,0.56,1.0,1.4 \}$. Each curve terminates at the value of the transition temperature corresponding to its $k_F a$ value: in decreasing order these are $T/T_F\approx\{0.50, 0.24,0.13,0.064\}$. The dashed line is the high-temperature asymptote $1/\tau_D \propto 1/\sqrt{T}$ predicted in Ref.~\cite{bruun2011}.}
\end{center}
\end{figure}

To obtain results for $1/\tau_D$ at constant density, we need to solve the non-interacting equation of state $n = (1/V) \sum_{{\bm k}} n_F(\xi({\bm k}))$, which yields the function $\mu(T,n)$. For low temperatures, $\mu$ is given by the familiar relation $\mu = \epsilon_F [1 - (\pi^2/12)( k_B T/ \epsilon_F)^2]$. Here, the Fermi energy $\epsilon_F = k_B T_F = \hbar^2 k_F^2/2 m$ with the Fermi wavenumber $k_F = (6 \pi^2 n)^{1/3}$. The transition temperature can then be determined as a function of the dimensionless interaction parameter $k_F a$. We find at unitarity ($-1/k_F a = 0$) $T_c \approx 0.50 T_F$ and that for weak coupling where ($-1/k_F a \gg 1$) $T_c \propto T_F \exp(-\pi/2 k_F |a|)$.
In Fig.~\ref{fig:BM} we show the dimensionless spin-transport relaxation rate $\hbar/\epsilon_F \tau_D$ obtained from Eq.~(\ref{eq:tauinv}) as a function of the reduced temperature $T/T_F$. Each curve terminates at the value of the transition temperature corresponding to its $k_F a$ value.
These curves illustrate a number of effects. Firstly, we see that, for all $k_F a$, the relaxation rate is enhanced (but remains finite) when approaching $T_c$ from above. This enhancement is larger in the strong-coupling case than in the weak-coupling case. The reason for this enhancement is that pairing fluctuations become important close to the superfluid transition, which lead to the appearance of a pole in $T^{MB}$ in the integrand of Eq.~(\ref{eq:tauinv}) at ${\bm K} =\mathbf{0}$ and $\omega =0$.

For large temperatures, all curves approach the dashed line in Fig.~\ref{fig:BM} given by $1/\tau_D = (\epsilon_F/\hbar)(32 \sqrt{2}/9 \pi^{3/2}) \sqrt{T_F/T}$ which falls off as $1/\sqrt{T}$, and was determined in \cite{bruun2011}. The $1/\sqrt{T}$ dependence can be understood by noting that $1/\tau \propto n v \sigma$, where $n$ is the density, $v$ the mean particle velocity, and $\sigma$ the cross section at unitarity. For temperatures $T \gg T_F$, the scattering cross-section is given by the square of the thermal de Broglie wavelength, and thus $\sigma$ decreases like $1/T$. Since for high temperatures it holds that $v \propto \sqrt{T}$, we find $1/\tau \propto 1/\sqrt{T}$. This high-temperature behavior was also measured experimentally \cite{sommer}.

In the weak-coupling case ($-1/k_F a \gtrsim 1$), we note an increase of $1/\tau_D$ with temperature up to $T\approx T_F$. An increase in temperature leads to a larger phase space available for scattering events due to reduced Pauli blocking. In the absence of many-body effects ($T^{MB} = V_0$), this effect would lead to the standard $T^2$ dependence of transport coefficients in a Fermi liquid. Indeed, when $T^{MB} = V_0$, we recover the weak-coupling result by Bruun, $1/\tau_D = (16 \pi/9)  (k^2_F a^2 \epsilon_F/\hbar) (T/T_F)^2$ for $T \ll T_c$ \cite{bruun2011}.

We note that our theory overestimates the $T_c$ at unitarity in comparison with the value $T_c = 0.15 T_F$ obtained by Monte Carlo and RG methods (see e.g. Ref.~\cite{RGTc}). In this sense, the experimental result for the spin-transport relaxation rate in Ref.~ \cite{sommer} should not be compared to the top curve in  Fig.~\ref{fig:BM} at unitarity, but instead to the second from below, which has a finite $k_F a$ value and $T_c = 0.13 T_F$. Then, we see that our result agrees qualitatively with the experimental results in Ref.~ \cite{sommer}.

\begin{figure}
\begin{center}
\includegraphics[width = 0.45 \textwidth]{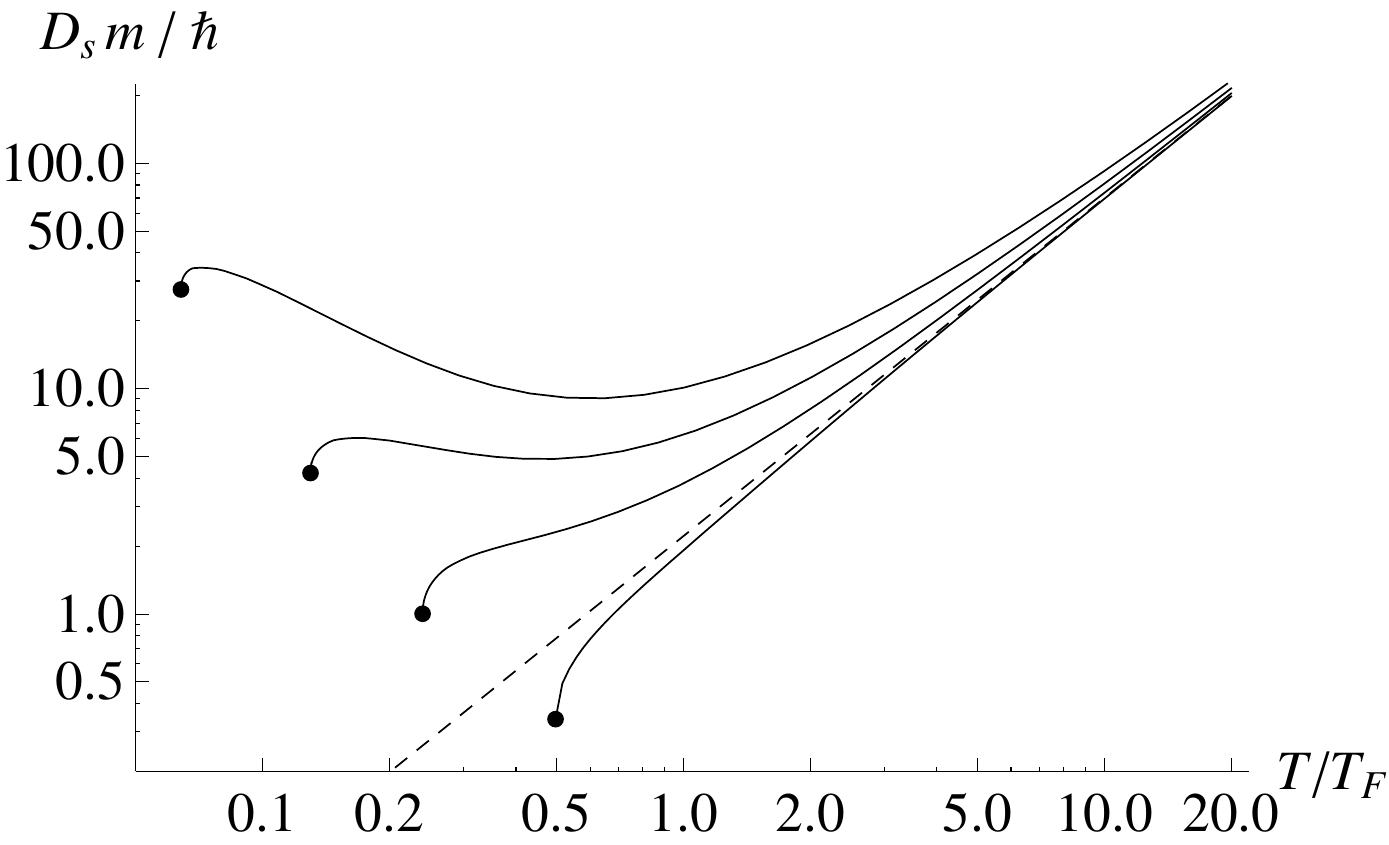}
\caption{\label{fig:BMD} The dimensionless spin-diffusion constant $D_s m /\hbar$ obtained from the result of Fig.~\ref{fig:BM} using the Einstein relation. From bottom to top the curves correspond to interaction parameters $-1/k_F a = \{0,0.56,1.0,1.4 \}$. Each curve terminates at the value of the transition temperature corresponding to its $k_F a$ value. The dashed line is the high-temperature asymptote $D_s =  (\hbar/m) (9 \pi^{3/2}/16 \sqrt{2}) (T/T_F)^{3/2}$}
\end{center}
\end{figure}

To obtain the diffusion constant $D_s$ we use the Einstein relation $D_s  = \sigma_D/\chi_s$. For the spin susceptibility, we use the non-interacting result $\chi_s =  - (1/V) \sum_{{\bm k}}  n'(\xi({\bm k}))$. The result is shown in Fig.~\ref{fig:BMD} where we plot the diffusion constant scaled by $m/\hbar$ versus the reduced temperature. The vertical order of the curves is reversed with respect to Fig.~\ref{fig:BM}. Each curve again terminates at the value of the transition temperature corresponding to its $k_F a$ value. The remarks made about Fig.~\ref{fig:BM} for a large part carry over to our results for the diffusion constant in Fig.~\ref{fig:BMD}. We see a suppression of $D_s$ when approaching the transition for all $k_F a$ values. For the top two curves with $-1/k_F a = \{1.0,1.4 \}$ we see that the non-monotonic behavior of $1/\tau_D$ is also present in the curves for $D_s$, which for these $k_F a$ values now have a minimum around $T/T_F = 0.5$. Since $\chi_s$ follows the Curie law $\chi_s = n/k_B T$ for high temperatures , we find that for high temperatures $D_s =  (\hbar/m) (9 \pi^{3/2}/16 \sqrt{2}) (T/T_F)^{3/2}$, which is shown as the dashed line in Fig.~\ref{fig:BMD}. We note the factor 2 difference with the result for this asymptote from Ref.~\cite{bruun2011}, which is caused by a different definition of $\sigma_D$. We note that if we again compare the second curve from above, which has a finite $k_F a$ value and $T_c = 0.13 T_F$, with the experimental results from Ref.~ \cite{sommer} at unitarity, we see qualitative agreement: both curves flatten off for temperatures below $T/T_F =0.5$. Considering our results, we see that this behavior is actually the crossover behavior between the monotonic behavior of the unitary (bottom) curve in Fig.~\ref{fig:BMD} and the non-monotonic behavior of the weak-coupling (top) curve. Clearly, a strength of our approach is that we can determine the whole temperature range using a single approach.
\section{Diagrammatic Theory}
\label{sec:dia}
In the second part of this work, we use a diagrammatic approach to determine the self-energy, spectral functions, and transport coefficients.
\subsection{Definitions}
First, we recall some aspects of the microscopic theory of an interacting gas of fermions. The central quantity in this theory is the self-energy, which is the sum of all one-particle reducible diagrams, and the spectral function derived from it. For reference we give the expressions for the spectral function, and the density of states and the particle density in an interacting system in terms of the spectral function. For a system which is translationally invariant in space and time, the retarded fermion Green's function is given by
\begin{equation}
G({\bm k}, \omega) = \frac{-\hbar}{- \hbar \omega^+ + \xi({\bm k})+\hbar \Sigma({\bm k},\omega)},
\end{equation}
where $\Sigma({\bm k},\omega)$ is the retarded self-energy. The spectral function describes the weight of a single-particle excitation and is given by
\begin{multline} \label{eq:rho}
\rho({\bm k}, \omega) = - \frac{1}{\pi \hbar} \Im\left[G({\bm k}, \omega)\right] = \\ - \frac{1}{\pi \hbar}
\frac{ \Im \Sigma({\bm k},\omega)}{[-  \omega + \xi({\bm k})/\hbar+ \Re \Sigma({\bm k}, \omega)]^2+ \Im^2 \Sigma({\bm k}, \omega)}.
\end{multline}
In the absence of interactions, $\Sigma =0$ and the spectral function is a $\delta$-peak at $\hbar \omega =\xi({\bm k})$. A nonzero real part of $\Sigma$ will shift the $\delta$-peak to the solution of $\omega =  \xi({\bm k})/\hbar+ \Re \Sigma({\bm k}, \omega)$, while a nonzero $\Im \Sigma$ will lead to a broadening of the peak, indicating a finite lifetime of the single-particle excitation. In particular, a constant $\Im \Sigma = -1/2 \tau$ will lead to a spectral function with a Lorentzian shape with width $\tau$. We note several relations for the spectral function that we use later. The sum rule states that
\begin{equation}
\int d \hbar \omega \rho({\bm k}, \omega) =1,
\end{equation}
and follows from the anticommutation relation of the fermion fields. The density of states is
\begin{equation}
\nu(\omega) = \frac{1}{(2 \pi)^3} \int d {\bm k} \rho({\bm k}, \omega),
\end{equation}
and the equation of state is
\begin{equation}
n = \int d \hbar \omega n_F(\omega) \nu(\omega).
\end{equation}
Using the Kubo formalism, we obtain the spin-transport relaxation rate $1/\tau_D$ and spin susceptibility $\chi_s = \partial (n_\uparrow - n_\downarrow)/\partial (\mu_\uparrow - \mu_\downarrow)$  in terms of the spectral function. These are in the absence of vertex corrections given by
\begin{equation} \label{eq:ktau}
\frac{1}{\tau_D} = - \frac{3 m n}{\pi \hbar^3} \left[\frac{1}{V} \sum_{{\bm k}} {\bm k}^2 \int d \hbar \omega  \rho^2({\bm k},\omega) n'(\hbar \omega)\right]^{-1},
\end{equation}
and
\begin{multline} \label{eq:kchi}
\chi_s  = - \frac{1}{V} \sum_{{\bm k}} \int d \hbar \omega d \hbar \omega' \\ \times \rho({\bm k},\omega) \rho({\bm k},\omega') \frac{n_F(\hbar \omega) - n_F(\hbar \omega')}{\hbar \omega-\hbar \omega'}.
\end{multline}
We note that Eq.~(\ref{eq:ktau}) can be derived by calculating the ordinary Drude conductivity of the atoms with spin up, say, in the presence of scattering from an `external' potential which is due to the presence of atoms with spin down. We note that for $\Sigma = - i/2 \tau$, Eq.~(\ref{eq:ktau}) gives $1/\tau_D = 1/\tau$, and that in the non-interacting case for low temperatures, Eq.~(\ref{eq:kchi}) reduces to the well known $\chi_0 = 3 n/2 \epsilon_F$. Finally, we note that we can obtain the spin-diffusion constant $D_s$ via the Einstein relation $D_s  = \sigma_D/\chi_s$.
\begin{center}
\begin{figure}
\includegraphics[width = 0.45 \textwidth]{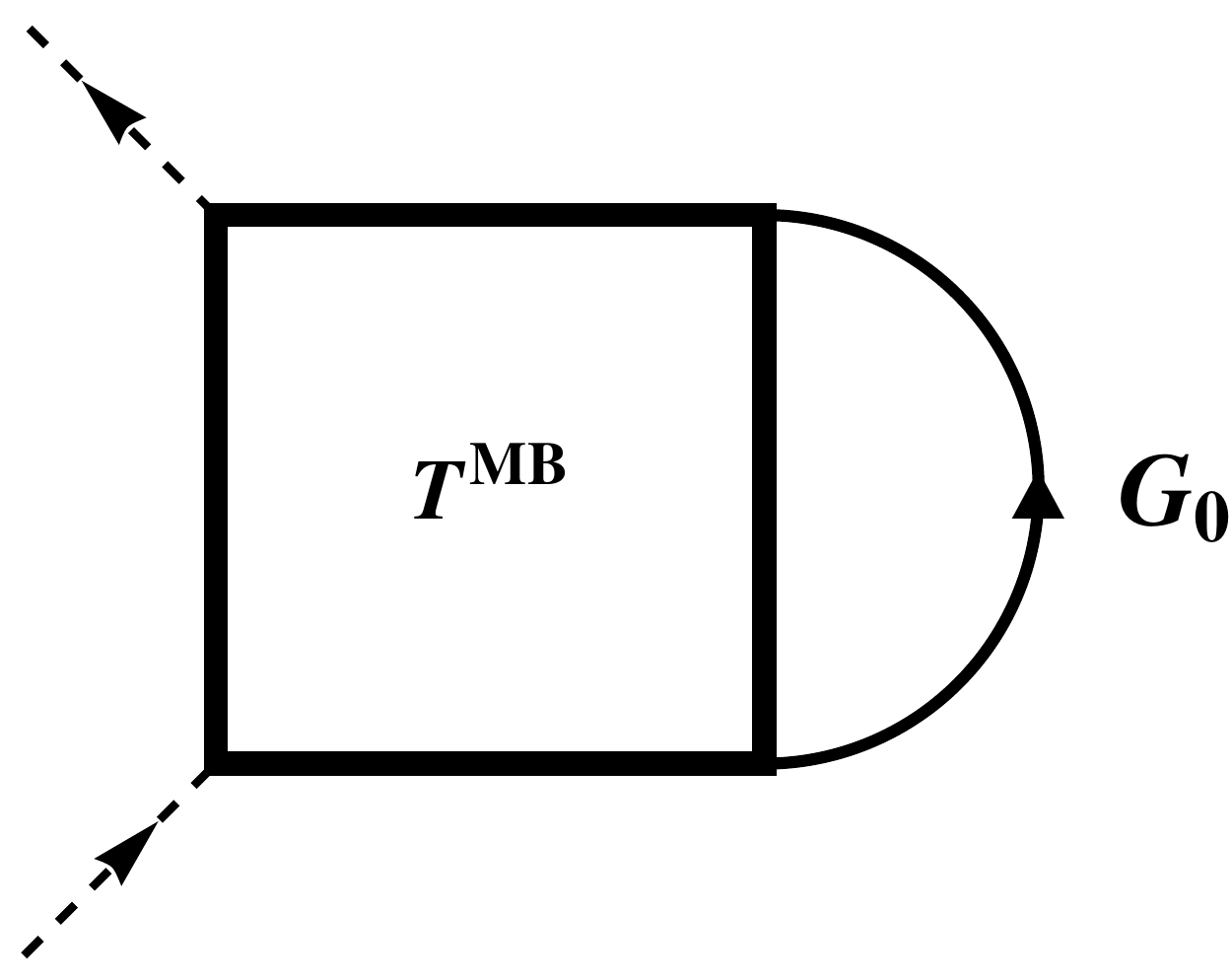}
\caption{\label{fig:dia} The self-energy $\Sigma $ is the many-body $T$-matrix closed with a single non-interacting Green's function line. The external lines are indicated by the dashed lines}
\end{figure}
\end{center}
\subsection{Self-energy}
We calculate the fermion self-energy which is appropriate for this system both close to the BCS pairing transition and for high temperatures at unitarity \cite{vctosi}. Indeed the many-body $T$-matrix closed with a non-interacting fermion line shown in Fig.~\ref{fig:dia}, incorporates the effect of enhancement of the interaction close to the BCS transition. In the dilute limit for weak interactions, where $T^{MB} \to V_0$, it reduces to the standard mean-field shift of the dispersion $V_0 n$. The expression for $\Sigma$ as a function of momentum ${\bm k}$ and fermionic Matsubara frequency $\omega_n$ is
\begin{multline} \label{eq:Si1}
\Sigma({\bm k},i \hbar \omega_n)= \frac{1}{\hbar^2 \beta} \frac{1}{V} \sum_{{\bm K},\omega_m} \\ \times T^{MB}({\bm K},i \hbar \omega_m)  G_0({\bm K}-{\bm k},i \hbar( \omega_m- \omega_n)),
\end{multline}
where the summation is over the bosonic Matsubara frequencies $\omega_m$ and $G_0$ is the non-interacting Green's function. We compute the Matsubara summation of Eq.~(\ref{eq:Si1}) to obtain
\begin{multline} \label{eq:Si2}
\Im \Sigma({\bm k},\hbar \omega)= \frac{1}{\hbar} \frac{1}{V} \sum_{{\bm K}}
[ n_F(\xi({\bm K}-{\bm k})) + n_b(\xi({\bm K}-{\bm k})+\hbar \omega)] \\ \times \Im[T^{MB}({\bm K},\xi({\bm K}-{\bm k}) + \hbar \omega)].
\end{multline}
Using $\Im[T^{MB}({\bm K},\hbar \omega)] = \Im[\Xi({\bm K},\hbar \omega)] |T^{MB}({\bm K},\hbar \omega)|^2$, we can rewrite Eq.~(\ref{eq:Si2}) as
\begin{multline} \label{eq:Si3}
\Im \Sigma({\bm k},\hbar \omega)= -\frac{\pi}{\hbar V^2} \sum_{{\bm k}_2,{\bm k}_3,{\bm k}_4}\\
\times \delta(\hbar \omega  + \xi({\bm k}_2) - \xi({\bm k}_3) - \xi({\bm k}_4)) \delta_{{\bm k}+{\bm k}_2, {\bm k}_3 + {\bm k}_4} \\ \times |T^{MB}({\bm k} + {\bm k}_2,\hbar \omega + \xi({\bm k}_2))|^2 \\ \times [n_2 (1-n_3)(1-n_4) + (1-n_2) n_3 n_4],
\end{multline}
where we again used the shorthand notation $n_i = n_F(\xi({\bm k}_i))$. Since $\Im \Sigma$ is a single-particle relaxation rate, we see that this rate is exactly the sum of the in and out rates for the state with momentum ${\bm k}$ and energy $\hbar \omega$ as would be obtained from a Fermi's golden rule expression. In Eq.~(\ref{eq:Si3}), we introduce the center-of-mass momentum ${\bm K} = {\bm k}+{\bm k}_2$, and an additional energy integral
\begin{multline} \label{deltas2}
\delta(\hbar \omega  + \xi({\bm k}_2) - \xi({\bm k}_3) - \xi({\bm k}_4)) \\ = \int d (\hbar \Omega) \delta(\hbar \omega  + \xi({\bm k}_2) - \hbar \Omega) \delta(\xi({\bm k}_3) + \xi({\bm k}_4) - \hbar\Omega).
\end{multline}
After some rewriting, we then arrive at the following expression for $\Im \Sigma$
\begin{multline} \label{eq:Si4}
n_F(\omega) (1-n_F(\omega) )\Im \Sigma({\bm k},\hbar \omega)= \frac{m}{8 \pi^2 k \hbar^3} \int_R d (\hbar \Omega)  d K\\
\times \frac{|T^{MB}(K,\hbar \Omega)|^2}{\sinh^2(\beta \hbar \Omega/2)}  \Im[\Xi(K,\hbar \Omega)]  \\
\times [1- n_F(\hbar \Omega -\hbar \omega)- n_F(\hbar\omega)],
\end{multline}
which is a convenient starting point for numerical evaluation. In going from equation Eq.~(\ref{eq:Si3}) to Eq.~(\ref{eq:Si4}), we introduced $\Im \Xi$ from Eq.~(\ref{eq:xi}). The delta functions on the right hand side of Eq.~(\ref{deltas2}) restrict the integration range to $R$ which is defined by the following inequalities
\begin{equation}
2 \mu+\hbar \omega-\frac{1}{2} \epsilon(K)>0 \quad \text{and} \quad \mu + \hbar \Omega - \hbar \omega>0
\end{equation}
and
\begin{multline}
|\sqrt{\epsilon(k)} - \sqrt{\mu + \hbar \Omega - \hbar \omega}| <  \sqrt{\epsilon(K)} \\ < \sqrt{\epsilon(k)} + \sqrt{\mu + \hbar \Omega - \hbar \omega},
\end{multline}
where we defined $\epsilon(k) = \hbar^2 k^2 /2m$. Then, the real part of the self-energy is obtained by a Kramers-Kronig transform
\begin{equation} \label{eq:SiR}
\Re \Sigma({\bm k},\hbar \omega) = \frac{1}{\pi} \int d \omega' \frac{\Im \Sigma({\bm k},\hbar \omega)}{\omega' - \omega}.
\end{equation}
In the next subsection, we show the results for the spectral function and spin-transport coefficients obtained by evaluating Eqs.~(\ref{eq:Si4}) and (\ref{eq:SiR}).
\subsection{Results}
\begin{center}
\begin{figure}
\includegraphics[width = 0.45 \textwidth]{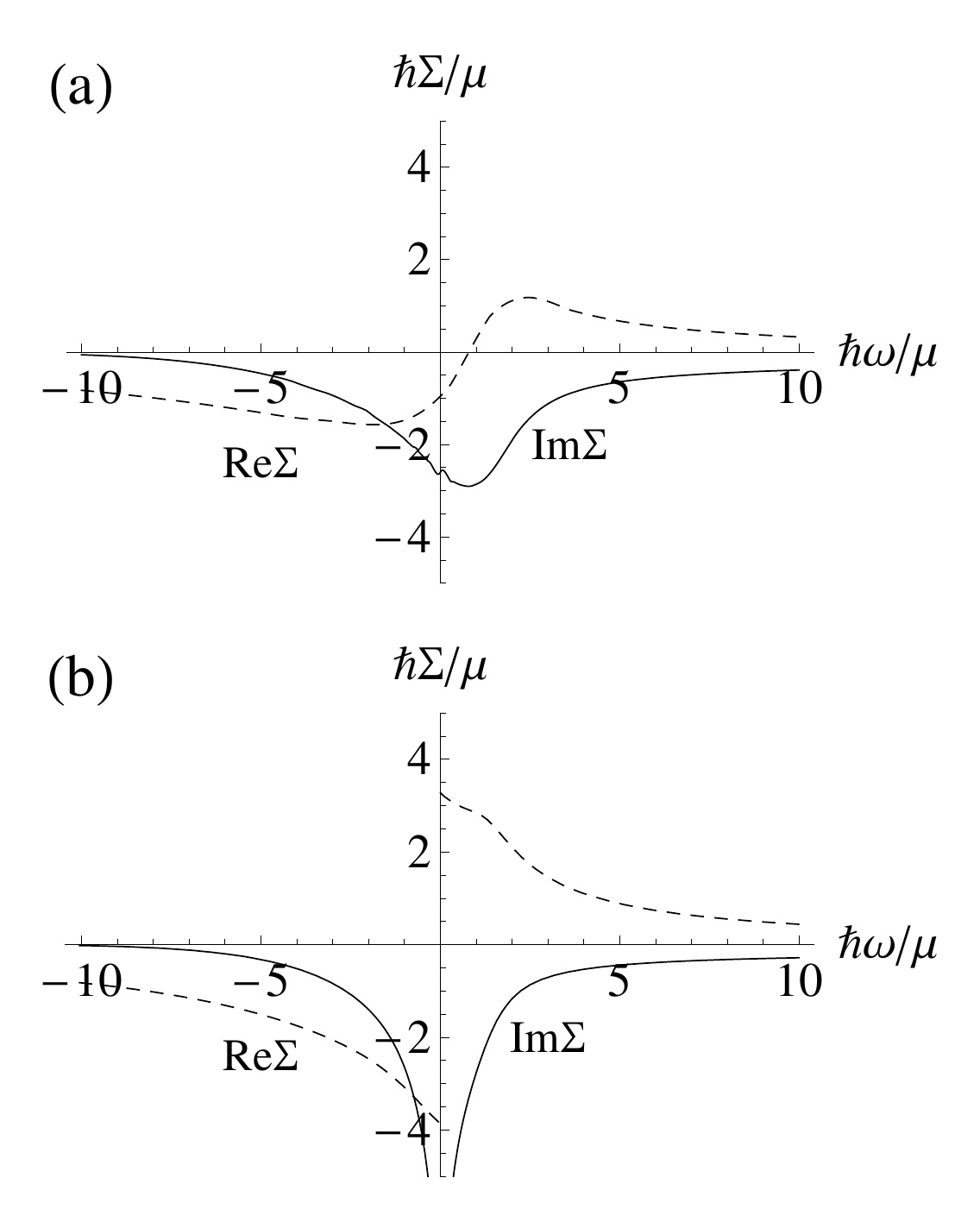}
\caption{\label{fig:parts} The imaginary (solid line) and real (dashed line) parts of the self-energy at unitarity for $\hbar k = \sqrt{ 2 m \mu}$ for $k_B T/\mu = 1$ (a) and $T = T_c$ (b).}
\end{figure}
\end{center}
\begin{center}
\begin{figure}
\includegraphics[width = 0.45 \textwidth]{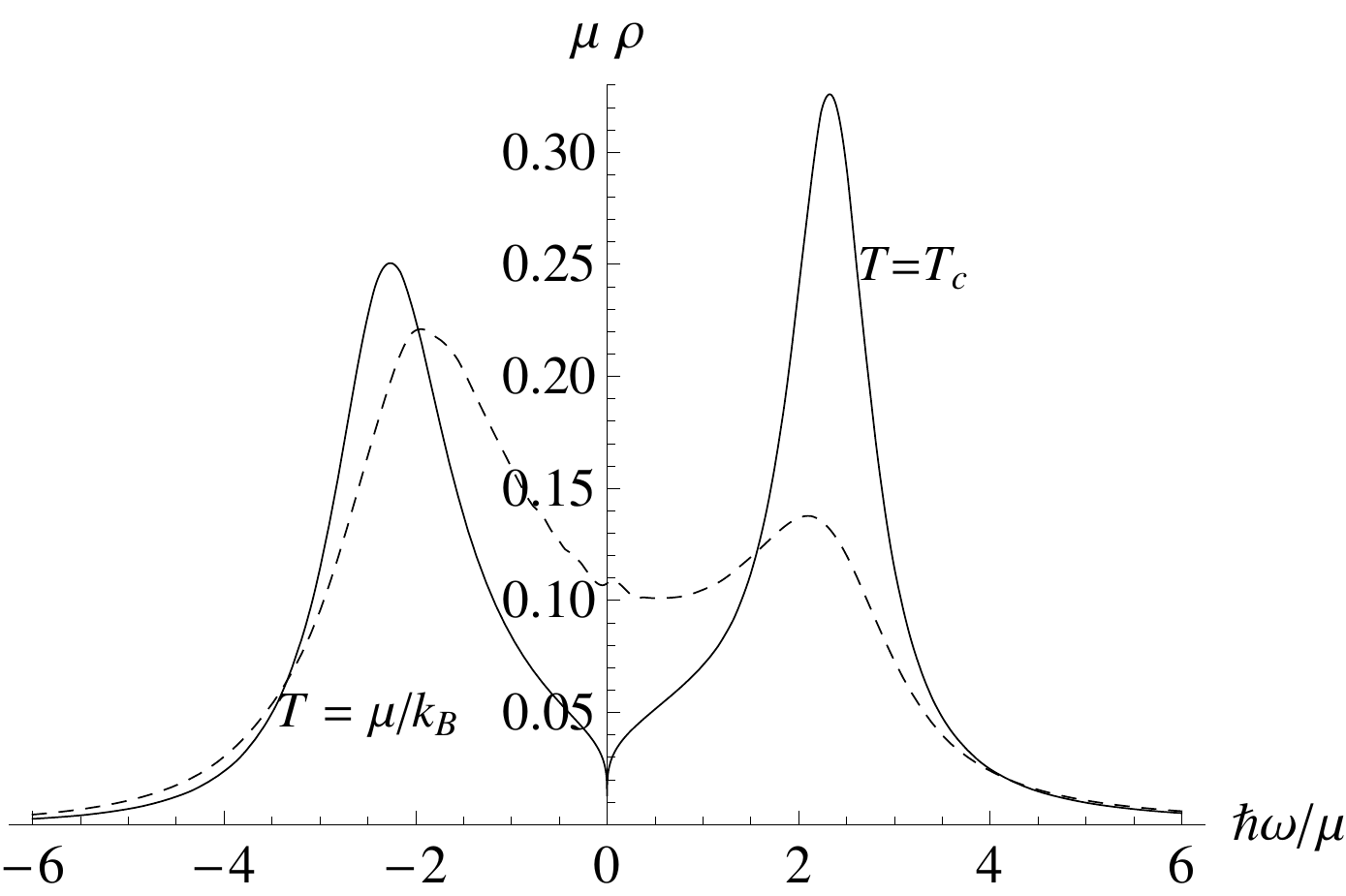}
\caption{\label{fig:rho} The spectral function $\rho$ at unitarity for $\hbar k = \sqrt{ 2 m \mu}$ for $k_B T/\mu = 1$ (dashed line) and $T = T_c$ (solid line).}
\end{figure}
\end{center}
By evaluating Eqs.~(\ref{eq:Si4}) and (\ref{eq:SiR}) numerically, we obtain Fig.~\ref{fig:parts} where we show the imaginary (solid line) and real (dashed line) parts of the self-energy for a typical momentum $\hbar k = \sqrt{ 2 m \mu}$ at unitarity.  Part (a) of Fig.~\ref{fig:parts} shows the results far from the transition at $ T = \mu/k_B \approx 1.5 T_c$, and part (b) of Fig.~\ref{fig:parts}) shows the results at the transition $T = T_c$. From Fig.~\ref{fig:parts} we see that upon approaching the transition, a divergence develops at $\omega =0$ in the imaginary part of the self-energy and a corresponding discontinuity in the real part. Closer inspection shows that the imaginary part diverges logarithmically. When $\Im \Sigma$ diverges as $x \log(\Delta \omega)$, with $\Delta \omega$ the distance from the pole, the resulting jump in the real part is $\pi x$. For general ${\bm k}$, the position of the divergence is $\hbar \omega = -\xi({\bm k})$, which can be understood as follows: the transition is accompanied by a pole in the many-body $T$-matrix $T^{MB}({\bm K},\hbar \Omega)$ for ${\bm K} ={\bf 0}$ and $\hbar \Omega = 0$. In the expression for $\Im\Sigma$ Eq.~(\ref{eq:Si3}) the $T$-matrix enters as $T^{MB}({\bm k} + {\bm k}_2,\hbar \omega + \xi({\bm k}_2))$, so that for $\hbar \omega = -\xi({\bm k})$, the conditions ${\bm K} ={\bf 0}$ and $\hbar \Omega = 0$ are satisfied when ${\bm k}_2 = -{\bm k}$.

As a check, we may also obtain the logarithmic divergence analytically. The divergence is caused by the development of a pole at $\Omega =0$ and $K = 0$ in Eq.~(\ref{eq:Si4}). For small $\Omega$ and $K$, $T^{MB} \propto 1/(a_1 \Omega + a_2 K^2 + a_3 (T-T_c)+i a_4 \Omega)$ (with the $a_i$'s real) while the rest of the integrand is linear in $K$. Integration over $\Omega$ and $K$ then yields a divergence $\log(T-T_c)$. In contrast, for small $\Omega$ and $K$ the integrand in the Bolzmann case Eq.~(\ref{eq:tauinv}) goes as $K^2$. Integration shows that there is then no divergence, as was shown in Fig.~\ref{fig:BM}.

From Eq.~(\ref{eq:rho}) we see that a divergence of $\Im \Sigma$ will lead to a suppression of the spectral function $\rho$. Indeed in Fig.~\ref{fig:rho} we show the spectral function $\rho$ at unitarity for $\hbar k = \sqrt{ 2 m \mu}$, for $k_B T/\mu = 1$ (dashed line) and $T = T_c$ (solid line). Already quite far from the transition at $T = \mu/k_B \approx 1.5 T_c$, interactions cause a suppression of the spectral weight close to $\omega=0$, which moves into peaks on both sides of $\omega=0$ in order to keep the sum rule satisfied. At $T = T_c$, this suppression is maximal, and here $\rho$ vanishes logarithmically as $|\omega| \to 0$.
\begin{center}
\begin{figure}
\includegraphics[width = 0.45 \textwidth]{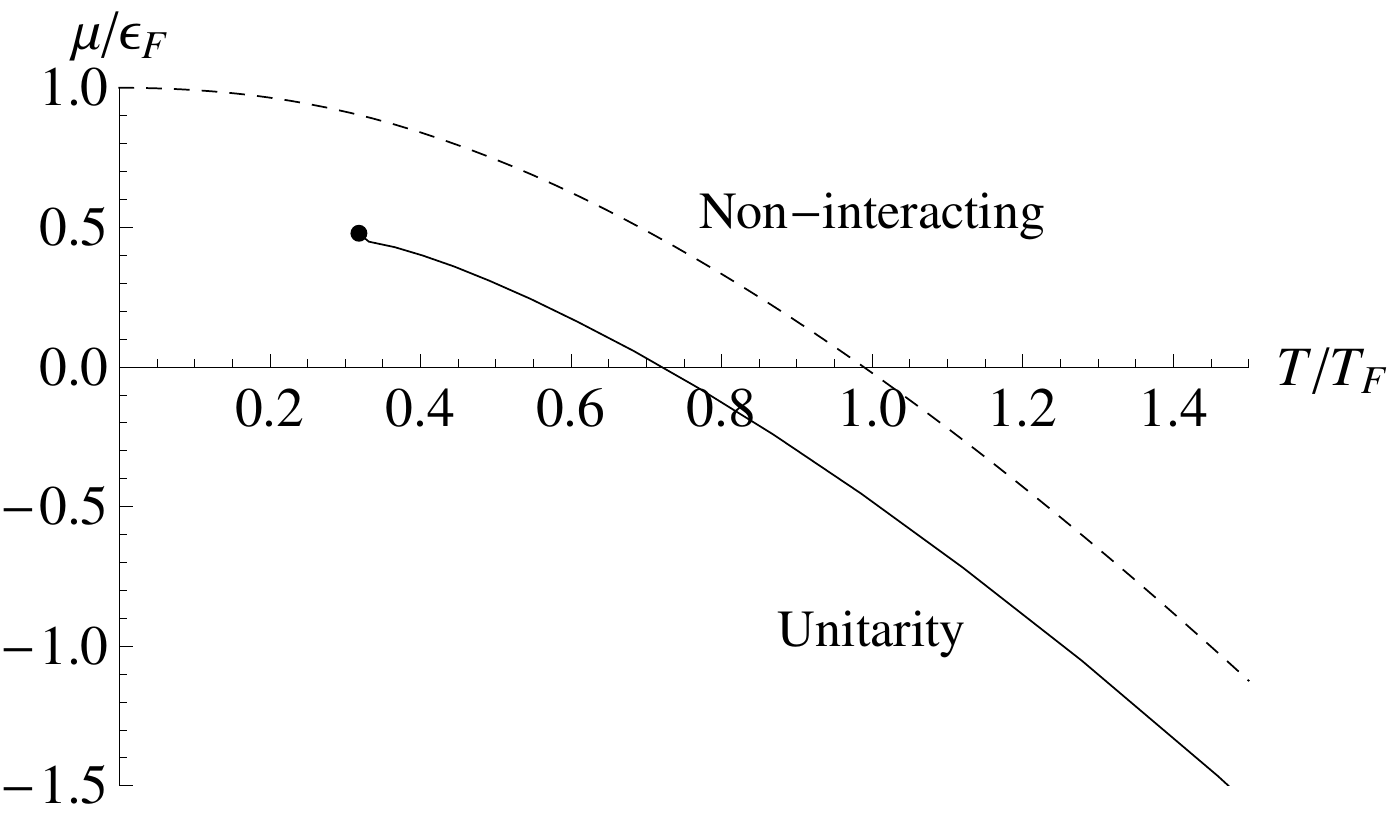}
\caption{\label{fig:num} The solution of the number equation giving $\mu$ as a function of temperature. The dashed line is the non interacting result. The solid line gives the result at unitarity.}
\end{figure}
\end{center}
\begin{center}
\begin{figure}
\includegraphics[width = 0.45 \textwidth]{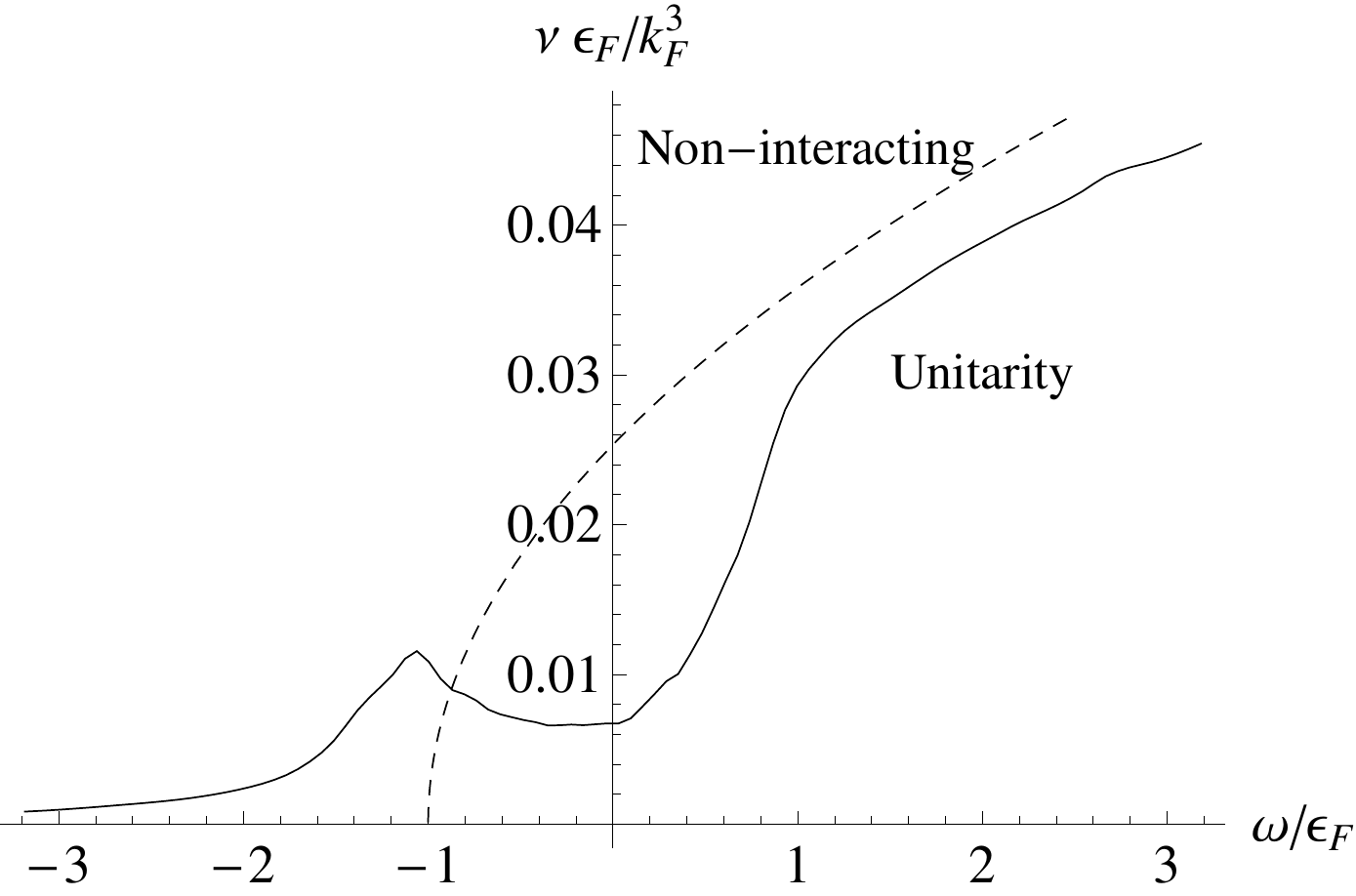}
\caption{\label{fig:DOS} The density of states $\nu$ versus $\omega$. The dashed curve is the non-interacting result for $T=0$, while the solid curve correspond to the unitary case for $T=T_c$.}
\end{figure}
\end{center}
In Fig.~\ref{fig:num} we show the solution of the equation of state $n = \int d (\hbar \omega) n_F(\omega) \nu(\omega)$ for $\mu$. The dashed line corresponds to the non-interacting result, which was used to obtain the Boltzmann results in Fig.~\ref{fig:BM}. The solid line is the solution of the number equation at unitarity, which terminates at the critical temperature $T_c \approx 0.29 T_F$. The difference of this value for $T_c$ with the mean-field value for $T_c$ at unitarity $T_c \approx 0.5 T_F$ used in Sec.~\ref{sec:BM} is that there, we used the non-interacting equation of state. That $\mu$ is lower than the non-interacting value is intuitively clear: an attractive interaction will lower the energy of the atoms and thus $\mu$ must also be lowered in order to keep the number of particles the same.

In Fig.~\ref{fig:DOS} we show the density of states $\nu$ versus $\omega$. The dashed curve is the non-interacting result at $T=0$, which goes as $\nu \propto \sqrt{\omega - \epsilon_F}$. The solid curve is the result at unitarity at $T = T_c$. We see that the suppressed spectral weight around $\omega = 0$ leads to a greatly reduced density of states, a so-called pseudogap. We note that since $T_c$ is not particularly low, the spectral functions are never very sharply peaked, and the density of states remains nonzero in the pseudogap in agreement with the results of Ref.~\cite{pseud}.
\begin{center}
\begin{figure}
\includegraphics[width = 0.45 \textwidth]{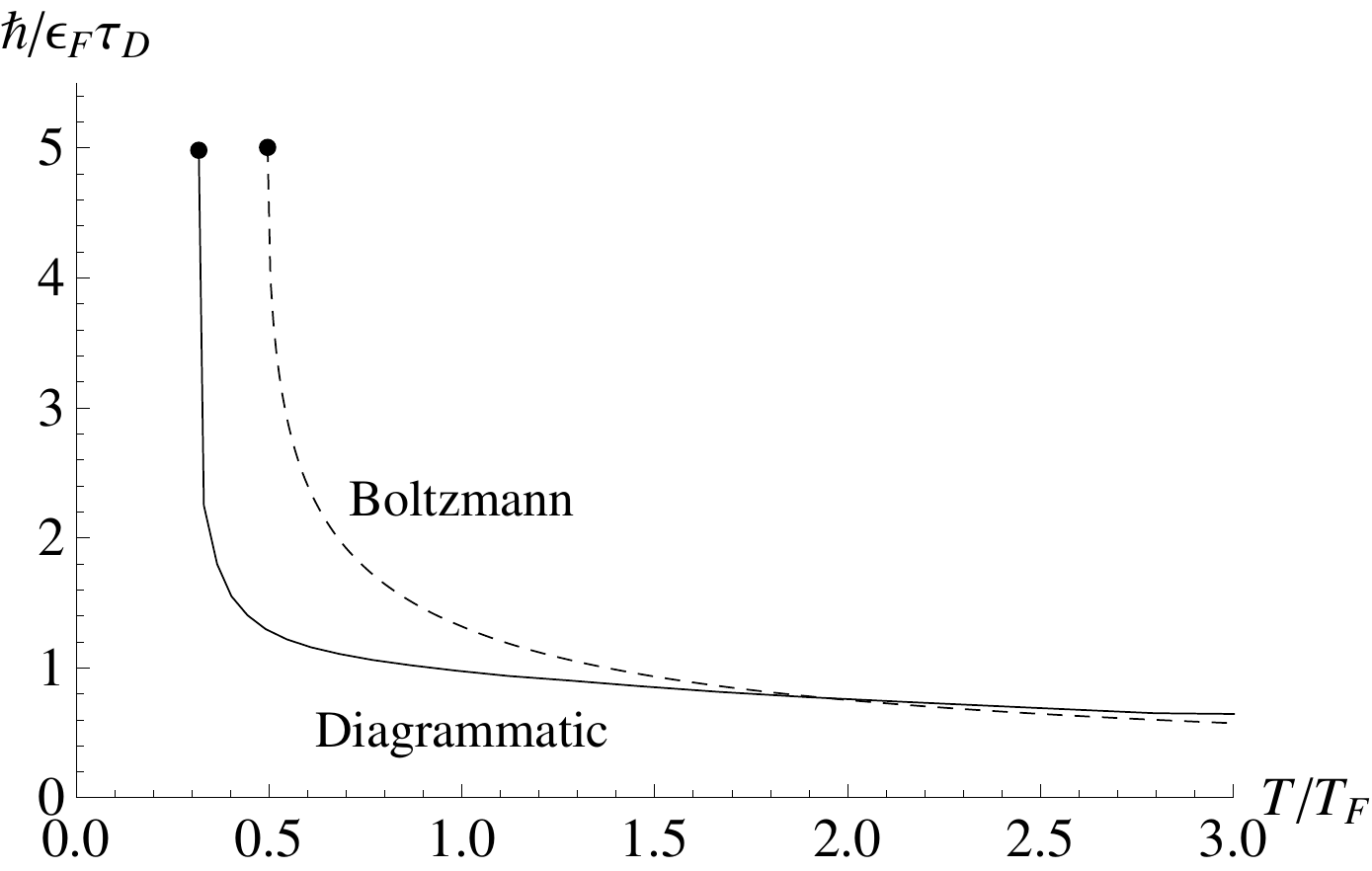}
\caption{\label{fig:tinv} The spin-transport relaxation rate $1/\tau_D$ as a function of temperature at unitarity. The dashed line is the Boltzmann result, the solid line the diagrammatic result obtained by evaluating Eq.~(\ref{eq:ktau}).}
\end{figure}
\end{center}
\begin{center}
\begin{figure}
\includegraphics[width = 0.45 \textwidth]{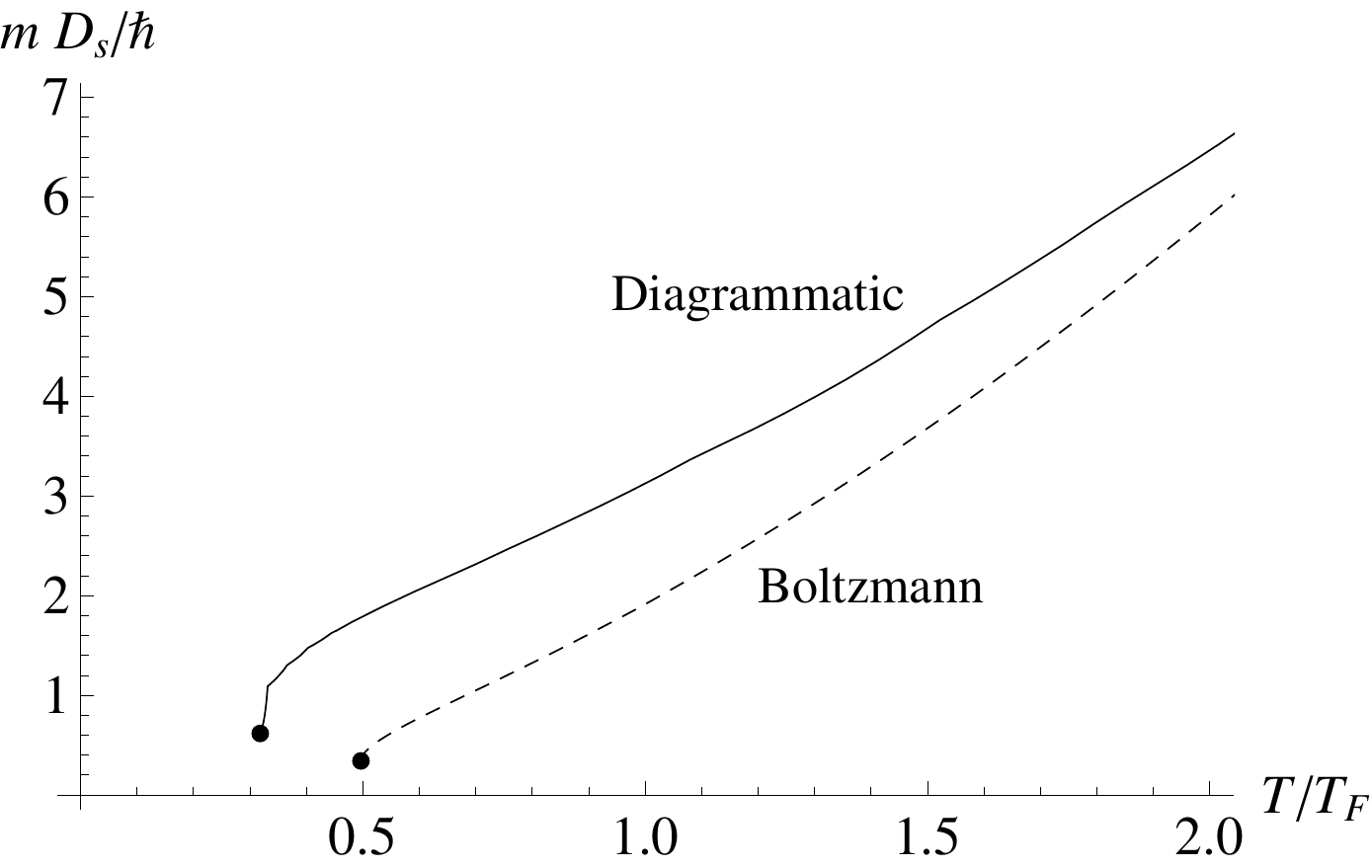}
\caption{\label{fig:Ds} The spin-diffusion constant $D_s$ as a function of temperature at unitarity. The dashed line is the Boltzmann result, the solid line the diagrammatic result obtained by evaluating the spin susceptibility in Eq.~(\ref{eq:kchi}) and using the Einstein relation.}
\end{figure}
\end{center}
In Fig.~\ref{fig:tinv} we show the spin-transport relaxation rate obtained by evaluating Eq.~(\ref{eq:ktau}) at unitarity as the solid line. The dashed line is the unitarity result from the Boltzmann calculation (the top line in Fig.~\ref{fig:BM}). Since both lines are determined using different equations of state, they terminate at different $T_c$'s. Similarly, in Fig.~\ref{fig:Ds} we show the spin-diffusion constant. The dashed line is the unitarity result from the Boltzmann calculation (the bottom line in Fig~\ref{fig:BMD}), while the solid line is obtained by evaluating the spin susceptibility in Eq.~(\ref{eq:kchi}) and using the Einstein relation to find $D_s$.

\begin{center}
\begin{figure}
\includegraphics[width = 0.45 \textwidth]{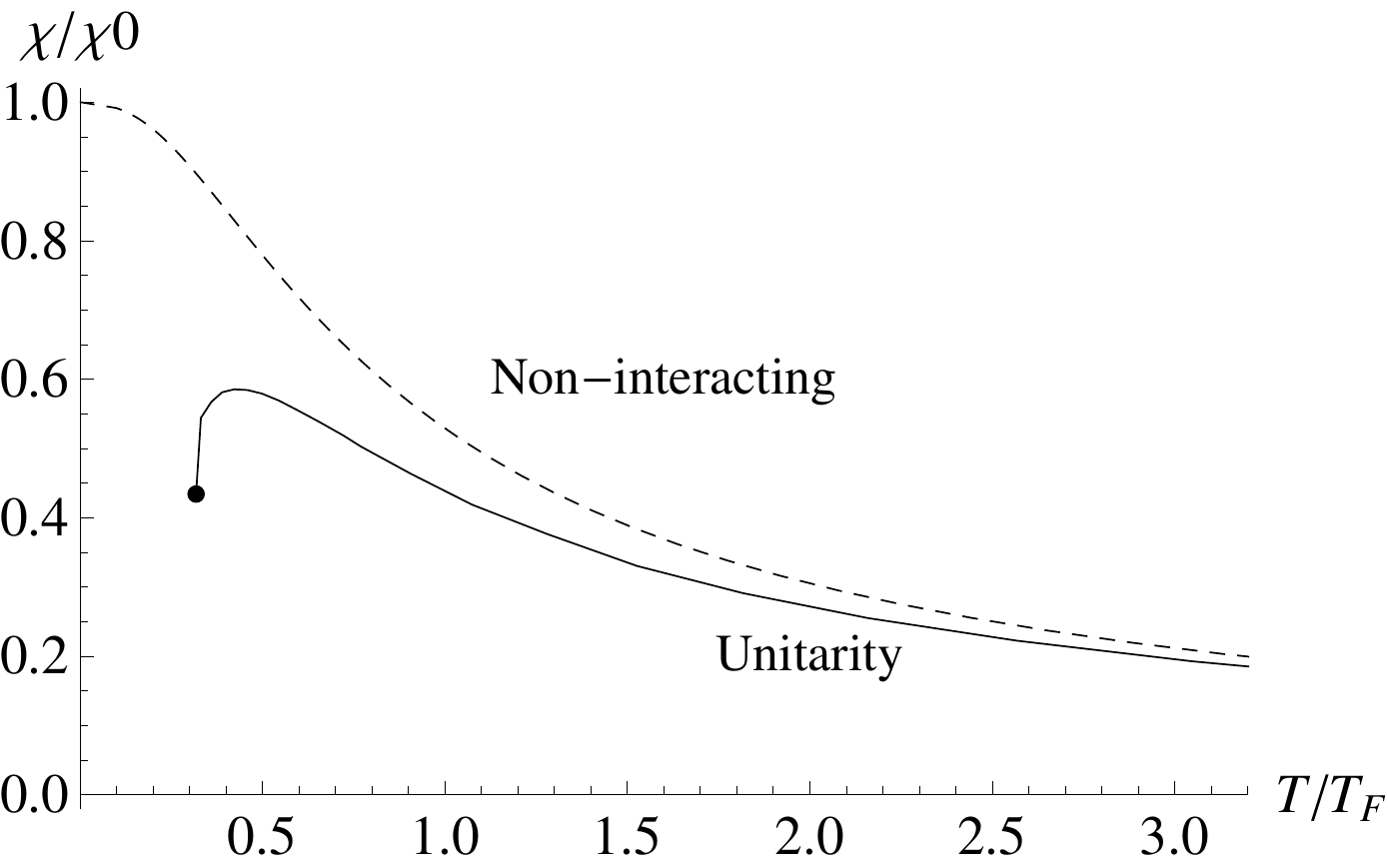}
\caption{\label{fig:chi} We show the spin susceptibility obtained by evaluating Eq.~(\ref{eq:kchi}) scaled by the non-interacting zero-temperature value $\chi_0 = 3 n/2 \epsilon_F$  as the solid line. The dashed line corresponds to the non-interacting result. The solid curve terminates at the critical temperature $T_c = 0.29 T_F$.}
\end{figure}
\end{center}

In Fig.~\ref{fig:chi} we show the spin susceptibility $\chi_s$ obtained by evaluating Eq.~(\ref{eq:kchi}) scaled by the non-interacting zero-temperature value $\chi_0 = 3 n/2 \epsilon_F$ as the solid line. The dashed line corresponds to the non-interacting result. We see that for large temperatures the lines converge, and that our result shows a downturn close to $T_c$. This downturn is expected physically, since the magnetic response should diminish, when $\uparrow$ and $\downarrow$ spins become more correlated \cite{theosuscep}. We note that in BCS theory the downturn of the spin susceptibility happens below $T_c$.
\section{Conclusions and discussion}
\label{sec:con}
We have determined the spin-diffusion constant and the spin-transport relaxation rate using a Boltzmann approach and a diagrammatic approach at unitarity, as shown in Figs.~\ref{fig:tinv} and \ref{fig:Ds}. On a diagrammatic level, the Boltzmann calculation takes into account vertex corrections, while our diagrammatic approach does not. Oppositely, the diagrammatic approach takes into account pseudogap physics which suppresses the spectral weight close to the Fermi level, while the Boltzmann approach does not. Surprisingly, we find qualitatively equivalent behavior for both transport coefficients using these approaches. Seemingly, pseudogap physics and vertex corrections are not of critical importance when evaluating these transport coefficients close to the BCS transition. To research this claim further, vertex corrections appropriate to the self-energy we used should be evaluated.

We note a recent preprint containing related work \cite{related}, in which Luttinger-Ward theory is used to study spin diffusion in Fermi gases.
\acknowledgments
This work was supported by the Stichting voor Fundamenteel Onderzoek der Materie (FOM), the Netherlands Organization for Scientific Research (NWO), and by the European Research Council (ERC). G.V. was supported by NSF Grant DMR -1104788. M.P. was supported by the Italian Ministry of Education, University, and Research (MIUR) through the program
``FIRB - Futuro in Ricerca 2010" - Project PLASMOGRAPH (Grant No. RBFR10M5BT).

\end{document}